# Uma abordagem Socio-Técnica de gerir a segurança de Informação: O uso do Artefacto *27001 Manager*

# A Socio-Technical approach to address the Information security: Using the 27001 Manager Artefact

Rui Shantilau[1] António Gonçalves[2], Anacleto Correia[3]

Integrity [1.]
Lisboa, Portugal
IPS[2]
Setúbal, Portugal
EN-CINAV[3]
Almada, Portugal

rs@integrity.pt
antonio.goncalves.pt@est.ips.pt.pt
cortez.correia@marinha.pt

**Resumo**. Em geral, a ótica cliente/fornecedor seguida pelas organizações, no que concerne à gestão da segurança da informação, assenta sobretudo na gestão de controles com base em normas tais como a ISO/IEC 27001:2015, resultando na produção de relatórios especialmente de análise técnica, em detrimento de uma abordagem sociotécnica. Isto conduz à perceção por parte do cliente da entrega de um produto em vez de um serviço. O produto em causa reduz-se a um conjunto de prescrições, por vezes não relacionadas, que se materializam numa visão descritiva e estática da gestão da segurança do cliente. Como resultado, o cliente dificilmente consegue utilizar o produto de forma continuada, acompanhando a dinâmica das alterações da sua organização, reconhecendo assim, valor na prestação efetuada pelo fornecedor.
A utilização do paradigma Lógica Dominante do Serviço (LDS), no desenvolvimento de uma oferta de gestão segurança da informação, auxilia a mudança de foco dos recursos tangíveis para os recursos intangíveis. Os aspetos de tangibilidade, materializado num documento que descreve as vulnerabilidades do cliente e os vetores de ataque, são remetidos para um plano secundário, face à relevância dos aspetos intangíveis, tal como a interação que se estabelece entre os especialistas do cliente e fornecedor.
Neste artigo propomos analisar sob a perspetiva de uma teoria sociotécnica, a Teoria da Atividade, o serviço fornecido por um artefacto designado por *27001 Manager*, destinado ao auxiliar todo o ciclo de análise, desenvolvimento e manutenção de um sistema de gestão de segurança de informação (SGSI). A analise pretende observar a interação existentes entre cliente/fornecedor, assumindo que o serviço é intrinsecamente dinâmico e intersubjetivo, i.e. fruto de um ajuste entre o cliente e o fornecedor.

**Palavras-chaves**: Teoria da Atividade, serviço, valor, co-criação, segurança de informação.

## 1 Introdução

Segundo vários estudos [1][2][3][4] a segurança da informação não é um produto ou uma tecnologia, mas sim uma abordagem sistémica que atua sobre três eixos - tecnologia, pessoas e processos -, tendo em conta um princípio basilar: a informação é um bem com valor que deve ser protegida.

As abordagens existentes, nomeadamente CIA[4], Parkerian Hexade[1] e IA[4], focam predominantemente os aspetos técnicos da segurança da informação, não definindo no contexto de uma organização, a importância do elemento humano na segurança da informação. Esta questão não se confunde com a utilização da engenharia social, como uma ferramenta explorada por possíveis atacantes.

Em geral o resultado do trabalho realizado no âmbito da segurança da informação, mais concretamente através da aplicação da norma ISO/IEC 27001:2015 é a entrega de um produto (i.e., relatório, plano de formação, desenvolvimento de um processo, etc.) que é o resultado, por vezes, de um trabalho sobretudo baseado no conhecimento tácito que os peritos tentam transmitir através de um relatório. Este produto descreve as vulnerabilidades encontradas e o impacto, para a proteção da informação, se essas vulnerabilidades forem exploradas por possíveis atacantes.

Uma solução de análise, desenvolvimento e implementação de um SGSI não pode somente contemplar os aspetos técnicos. Deve também considerar a relação de cooperação, coordenação e colaboração entre os intervenientes a qual é uma atividade complexa. O resultado esperado deve ser a criação de conhecimento organizacional para que se possa lidar com os desafios atuais e futuros relativos a gestão do SGSI. A dificuldade reside em capturar e partilhar entre os intervenientes o conhecimento organizacional, de forma coerente, compreensiva, consistente e concisa. O resultado do trabalho realizado, no âmbito do SGSI, deve constituir mesmo, um fator capaz de promover a compreensão partilhada na organização.

Neste artigo pretende-se contribuir para a clarificação dos benefícios do uso de um artefacto designado por *Manager 27001* para o suporte a um SGSI como uma ferramenta organizacional. Deste modo é considerada:
1. A abordagem da Lógica Dominante de Serviço, que será usada para descrever a colaboração entre os intervenientes do serviço prestado com o artefacto;
2. A apresentação de um modelo, segundo a Teoria da Atividade, que descreve genericamente um serviço, segundo a Lógica Dominante de Serviço;
3. A instanciação do modelo proposto em 2 para descrever o artefacto, na atividade de suporte ao SGSI.

O artigo está estruturado do seguinte modo: Na secção 2 são apresentados os principais conceitos da Lógica Dominante de Serviço. Na secção 3 são descritos os passos associados aos testes de intrusão. Em 4 é apresentado as vantagens e os problemas relacionados com a abordagem clássica dos testes de intrusão. Em 5 é feita uma descrição da Teoria da Atividade e os princípios basilares. Em 6 é descrito de que modo podemos utilizar a Teoria da Atividade para definir um serviço. Em 7 é mapeado este modelo para um caso de estudo realizado sobre testes de intrusão. Em 8 são apresentados trabalhos relacionados com a segurança da informação. Em 9 são apresentadas as conclusões e o trabalho futuro.

## 2 Lógica Dominante de Serviço

A distinção entre produto e serviço resulta do enfoque particular numa das seguintes abordagens distintas: a Lógica Dominante de Produto (LDP) e a Lógica Dominante de Serviço (LDS) [5]. O que distingue um serviço de um produto é o grau de importância dos elementos tangíveis (e.g., um software, servidor, router, etc.) dos intangíveis (e.g., competência, regras, conhecimento). Se o valor essencial é mais tangível do que intangível, este é considerado um produto. Se o valor essencial é mais intangível do que tangível, é um serviço. Contudo, nem sempre, esta distinção é tão clara, uma vez que existem produtos e serviços, constituídos por elementos tangíveis e intangíveis que contribuem para a entrega final ao cliente[5][6]. Na figura 1 é apresentado o modelo o LDS na linguagem de modelação baseada em factos (ORM) [7].

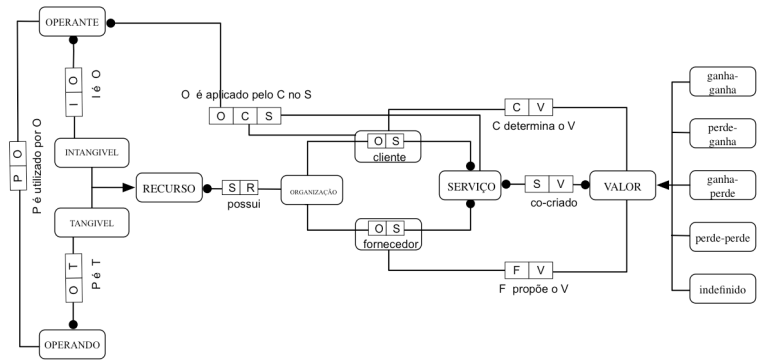

**Figura 1. Modelo Lógica Dominante de Serviço.**

Segundo o modelo apresentado na figura 1 podemos interpretar o serviço como sendo um acordo, em que um fornecedor utiliza os seus recursos operandos (tangíveis) e operantes (intangíveis) em benefício de um cliente. Desta iteração é possível descrever um conjunto de estados agrupados num modelo, designado por ISPAR (*Intercat-Serve-Propose-Agree-Realize*), que abrange todas as possíveis interações, incluindo tanto "sucesso" como "insucesso". Por exemplo, no percurso para o "sucesso", possui um conjunto de passos que ocorre, primeiro quando um serviço é proposto, ou seja, uma "proposta de valor" é comunicada. A comunicação requer um sistema entre o cliente e o fornecedor, de modo que ambos os interlocutores estejam envolvidos na interação. Em seguida o serviço é consumido. No caso de sucesso é reconhecido o valor do serviço prestado. Entre o sucesso e o insucesso, existe um conjunto finito de estados, que são descritos no modelo ISPAR[5].

Segundo a LDS o serviço representa mais do que o suporte a um produto. A co-criação existe quando há uma participação ativa do cliente, ou seja é gerado valor no serviço devido a sua configuração acompanhar as necessidades do cliente [5].

## 3   Sistema de Segurança de Informação

A partir de novembro de 2005, surgiu uma nova versão da norma de segurança de informação suportada no comité SC27. O objetivo desta norma é a segurança da cadeia de valor de uma organização contra os riscos de segurança da sua informação. Este novo padrão, designado por ISO/IEC 27001:2005 é um desenvolvimento anterior que por sua vez é uma evolução da norma BS7799-2 [8, 9].

O ISO/IEC 27001:2005 postula uma nova conceção de um sistema de gestão de informação. O conceito central foca-se no facto de que este sistema de gestão é orientada ao risco do negócio de uma organização, numa relação direta com a sua cadeia de valor.

Num SGSI, de acordo com a norma ISO 27001, somente os processos que têm uma contribuição crítica para o sucesso do negócio da organização são selecionados. Estes processos podem também ser resumidos sob o conceito de cadeia de valor [10]. Estes processos críticos da cadeia de valor são submetidas a uma análise de risco de segurança através de uma apreciação discricionária com base num conjunto vasto de controlos fornecidos pela norma. Mais concretamente 133 controlos.

Um SGSI [11] é um processo complexo de analisar continuamente impacto no negócio através de falhas na segurança de informação de uma organização. Constitui um modo relevante de avaliar a segurança das infraestruturas tecnológicas, aplicacionais e humanas, através de uma abordagem analítica para descobrir as vulnerabilidades e ameaças que podem comprometer a segurança da organização [3]. O objetivo dos SGSI é mitigar brechas, num ambiente controlado, no acesso aos sistemas de informação, de modo a proteger de possíveis práticas utilizadas por atacantes com intensões maliciosas [12].

Na tabela 1, estão resumidas as principais características de um SGSI com base em [13].

**Tabela 1. Principais características de um SGSI.**

| Foco | Benefício |
|---|---|
| Política de Segurança | Permite a uma organização expressar a sua intenção de proteger a informação, indica direções para a gestão e proporciona a outros informa os outros *stakeholders* na primazia do seu de esforço. |
| Comunicação e gestão da operação | Define a política de segurança na organização, em particular na redução do risco e analise o esforço de verificação dos controles e distribuição de responsabilidades através de procedimentos bem definidos. |
| Controlo de Acesso | Possibilita a definição de uma autoridade de controlo de acesso a recursos quer físicos que intangíveis. |
| Contexto de sistemas de informação | Definição de um processo integrado define o contexto dos sistemas de informação, incluindo a sua aquisição, desenvolvimento e manutenção. |
| Gestão de ativos | Define a identificação, controlo, classificação e atribuição de propriedade dos ativos relevantes para garantir a sua segurança. |

## 4 Descrição do Problema

O modelo clássico de suporte ao desenvolvimento de SGSI é enquadrado numa lógica dominante de produto que passa pela contratação, tipicamente de um conjunto de horas/dias de consultor, para verificar e propor os aspetos relacionados com a segurança de uma organização.

O modelo clássico possui diversos benefícios, nomeadamente é realizado por equipas especializadas (internas ou externas), favorece a deteção de vulnerabilidades, deteta falhas não convencionais ou triviais, possibilita contornar as ferramentas de proteção e problemas de arquitetura de segurança. O resultado pode, de acordo com as evidências de intrusão obtidas, promover a consciencialização da organização, o que poderá, inclusive, conduzir a um maior investimento na segurança.

Possui contudo alguns problemas, nomeadamente dificulta a abordagem segundo uma perspetiva de eficácia e melhoria contínua, porque não permite a monitorização constante da organização, uma vez que não acompanha o ciclo de mudança organizacional, porque é feito em períodos temporais com uma duração predefinida. Por outro lado, devido à limitação do tempo disponibilizado para a realização da implementação do SGSI, não permite aos especialistas aprofundar o seu conhecimento sobre a organização sobretudo sobre o negócio da organização e por isto dificulta a descoberta dos vetores de ataque às vulnerabilidades associados aos aspetos do negócio, pessoas e processos.

Apesar de para a vertente de gestão de risco o mercado oferecer já um considerável leque de soluções, em todas elas identifica-se imediatamente uma lacuna: a capacidade de identificação de qual a metodologia de análise de risco mais adequada à realidade da empresa. Neste sentido, é de realçar que há uma elevada quantidade de metodologias de análise de riscos, as quais apresentam desempenhos diferentes consoante o tipo de negócio ou as tecnologias associadas, entre muitos outros fatores. Como tal, a sua escolha não é linear.

Esta limitação encontra-se intimamente ligada à principal limitação das restantes soluções que o mercado oferece também para suporte à implementação e gestão de um SGSI: a capacidade de análise da situação, processos e requisitos existentes numa organização, e consequente definição do plano de ação.

Uma vez terminado a solução do SGSI e entregue o relatório, rapidamente este fica desatualizado, por um lado porque constantemente surgem um elevado número de novas vulnerabilidades as quais não foram consideradas. Isso faz com que o risco aumente assim que é finalizado o trabalho do consultor.

As soluções existentes, para além de invariavelmente serem direcionadas para a componente de gestão de riscos, não apresentam a capacidade de análise e definição de metodologias. Mesmo as soluções mais abrangentes, estão sempre dependentes de uma complexa fase de levantamento de requisitos, os quais impõem tanto uma elevada intervenção de pessoal técnico especializado como de técnicos da própria empresa. Com base nesse complexo e exaustivo trabalho é que se avança para a definição, parametrização ou mesmo desenvolvimento da ferramenta, ou ferramentas, que irão suportar o SGSI. Como tal, é um processo não só demasiado longo e de difícil execução, como também extremamente dependente de intervenção humana especializada e, como tal, altamente sujeito à existência de falhas, lacunas e limitações, fatos altamente contrários ao objetivo das soluções SGSI. De realçar ainda que estas dificuldades e limitações, têm impacto nas três fases do SGSI: conceção, implementação e ainda na operação. Tal pode levar a que apesar de se conseguir implementar um SGSI que cumpra inicialmente a norma, possa facilmente vir a não conseguir manter essa certificação.

Na relação entre quem realiza o trabalho de conceção do SGSI e a que se destina o SGSI existem diversos problemas. No modelo clássico o processo acaba por não ser claro o envolvimento do cliente no processo. Isto é, a participação do cliente está longe de fazer parte do processo. A grande maioria dos especialistas reconhece a relevância e a importância do envolvimento do cliente no processo, mas utiliza uma diversidade de argumentos que justificam a inconsistência entre as suas conceções e as suas práticas (e.g., falta de formação, necessidade de cumprir o trabalho contratualizado, proteção do conhecimento, etc.).

A avaliação do trabalho realizado acaba por ser fundamentalmente, quase uma autoavaliação do especialista. Normalmente não existe uma partilha real dos métodos e vetores de ataque realizados pelo especialista com o cliente. A avaliação do processo realizado no desenvolvimento do SGSI, é pois um processo pouco transparente. Os critérios de exploração, de captura de evidência e de classificação não são, em geral, explicitados nem clarificados ao cliente. Por fim, não se tem em conta a diversidade e a multiplicidade inerentes às atividades humanas, e a historicidade e o desenvolvimento da organização que por vezes dificulta ao especialista o conhecimento do modo como a organização é operada. Deste modo, a passagem do e o seu reconhecimento pela organização constituí uma tarefa complexa e por vezes resulta na insatisfação de ambas as partes. Por parte do especialista a falta do reconhecimento do trabalho desenvolvido; pela organização da relevância organizacional do esforço financeiro, face aos resultados alcançados.

Podemos concluir que o foco, do modelo clássico, atende às necessidades atuais e não futuras. Deste modo, não contribui por criar e manter um ambiente organizacional no cliente, no qual exista um envolvimento contínuo no propósito de promover o aperfeiçoamento continuado da segurança da informação na organização. Verifica-se ainda que os conceitos lineares de causa e efeito revelam-se insatisfatórios no processo de compreensão do comportamento de risco dos clientes, devido a factos sociais descritos por uma complexidade de múltiplos elementos (pessoas, ferramentas, regras, divisão de trabalho, etc.) que interagem sistemicamente uns com os outros numa organização.

## 5 Teoria da Atividade

A Teoria da Atividade é um enquadramento multidisciplinar que descreve como as pessoas interagem coletivamente, em contextos, para realizarem o seu trabalho. A cada contexto dá-se o nome de atividade. As atividades resultam de um processo de desenvolvimento socio-histórico, que descreve a estabilidade numa organização como uma exceção e as perturbações e conflitos como regra, agente de mudança e de inovação [14, 15].

A Teoria da Atividade faz uma clara distinção do papel das pessoas e das ferramentas. As pessoas executam, ao longo do tempo, as suas tarefas, conscientes (ações) e inconscientes (operações), com o auxílio das ferramentas, sendo que estes não são o objeto de interesse, mas sim o meio pelo qual é possível atingir-se o resultado pretendido [14].

Engeström [14] propôs um modelo sistémico para representar a estrutura da atividade (figura 4) e a descrição dos seus elementos (tabela 2).

Ele mostra graficamente as relações entre os elementos que constituem a atividade. Neste diagrama, o primeiro foco da análise do sistema de atividade é o ponto médio da face direita do triângulo (a produção de algum resultado). Na produção de qualquer Atividade, temos um sujeito, o objeto da atividade, as ferramentas usadas e as ações e operações que afetam o resultado [14].

Associado a teoria da Atividade existe um conjunto de princípios que são apresentados na tabela 3.

**Tabela 3. Princípios da Teoria da Atividade.**

| Principio | Descrição |
|---|---|
| Orientação a objetos | Reflete a natureza da atividade humana, a qual permite o controlo do comportamento com vista à satisfação de objetivos identificados, ou seja deve existir uma motivação associada a ação humana. |
| História e desenvolvimento | Reflete a evolução ao longo do tempo do trabalho realizado numa atividade e que estão incorporados nos elementos, nomeadamente na divisão de trabalho, nas regras e nas ferramentas |
| Artefactos | Estes só podem ser compreendidos no contexto da atividade, isto é a forma como são utilizados estão relacionados com a história da sua utilização ao longo do tempo. |
| Mediação | A grande maioria das interações do sujeito com o ambiente (pessoas e objetos) é mediada por artefactos. Artefactos podem ser tangíveis ou não. Um dos principais aspeto do conceito de mediação é que o uso dos artefactos acaba por transformar não apenas o objeto da ação, mas também o seu sujeito, uma vez que condicionam o modo como o sujeito interage. |
| Relação Sujeito-Sujeito | Representa o lado social e comunicativo de uma atividade que regulam a interação as pessoas e o espaço que cabe a cada um. |
| Contradições | São tensões acumuladas historicamente e que geram perturbações e conflitos mas também é o meio pelo qual a atividade pode evoluir. |
| multivocalidade | Representa os pontos de vistas distintos. A multivocalidade multiplica-se quando se considera uma rede de sistema de atividades (figura 5). |

Os princípios não são conceitos isolados, uma vez que cada um deles estão relacionados, logo interdependentes.

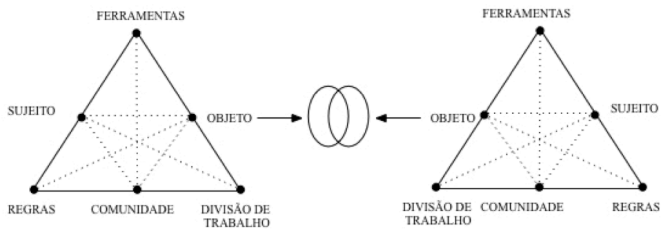

**Figura 3.** Rede de Sistema de Atividades

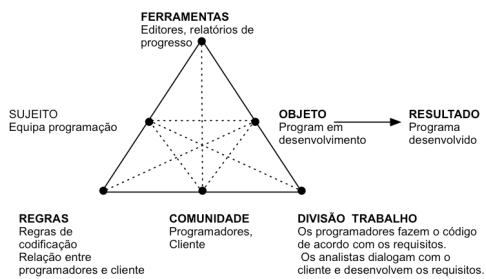

**Figura 2.** Sistema de Atividade

**Tabela 2. Descrição dos elementos**

| Item | Descrição |
|---|---|
| Sujeito | Ponto de vista de um indivíduo ou grupo. |
| Ferramenta | São recursos culturais ou instrumentos utilizados. |
| Regras | Normas, convenções que regulam a operação. |
| Comunidade | Indivíduos que partilham o mesmo interesse. |
| Divisão de trabalho | Distribuição de tarefas. |
| Objeto | Representa o espaço do problema |
| Resultado | Traduz-se numa alteração da realidade. |

## 5.1 Análise de Contradições Segundo a Teoria da Atividade

Segundo Engeström, as contradições constituem um elemento chave na teoria da atividade e são características dos sistemas de atividade. As contradições não são o mesmo que conflitos, mas tensões estruturais acumuladas historicamente dentro e entre sistemas que geram problemas, falhas e conflitos, mas também tentativas de inovação para mudar a atividade [15] (ver figura 6).

As ***contradições de primeira ordem*** refletem o conflito interno entre o valor de troca e o valor de uso, em cada canto ou vértice do triângulo da atividade (indicado com os círculos com o número 1 na figura 6). Estas contradições correspondem a tensões encontradas num elemento interno de uma dada atividade.

As ***contradições de segunda ordem*** estão entre os cantos do triângulo e ocorrem entre os elementos do sistema de atividade (indicado com os círculos com o número 2 na figura 6). Estas contradições aparecem quando um fator novo surge num dos elementos do sistema. Nestes casos, a manifestação da contradição não pode ser isolada e está relacionada com a relação entre dois ou mais elementos da atividade.

As ***contradições de terceira ordem*** ocorrem quando é introduzido o objeto e motivo de um outro sistema de atividade, culturalmente mais avançado, no sistema de atividade vigente (indicado com os círculos com o número 3 na figura 4). A manifestação deste tipo de contradições surge quando os conflitos podem limitar o desenvolvimento da atividade atual em relação a uma atividade hipotética culturalmente mais desenvolvida.

Finalmente, as ***contradições de quarta ordem*** ocorrem entre o sistema de atividade central e os sistemas de atividades circunvizinhas na rede de sistemas e emergem da interação da atividade central com as atividades periféricas (na figura 6 é referido com os círculos com o número 4). As atividades circunvizinhas incluem: (1) as atividades nas quais os objetos e produtos (ou resultados) da atividade central estão fixados (atividades/objeto); (2) as atividades que produzem as ferramentas chave para a atividade central (atividades de produção de ferramentas); (3) as atividades como aprendizagem e escolarização do sujeito (atividades de produção do sujeito); (4) as atividades de administração e legislação (atividades de produção de regras). A maioria das tensões ocorre nesta situação, onde normalmente uma dada atividade fica dependente de um resultado construído por outra.

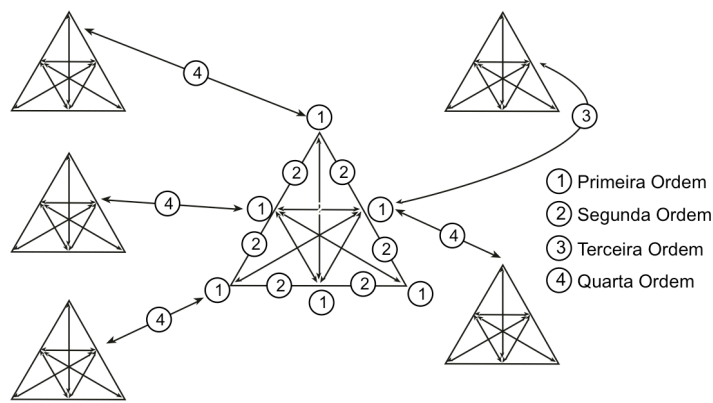

**Figura 4. Níveis de contradições**

## 5.2 Ciclo de aprendizagem expansiva

A partir da análise de uma rede de sistema de atividades é possível analisar as manifestações das contradições. A ascensão do abstrato para o concreto é alcançada pelas ações de aprendizagem que, juntas, formam um ciclo expansivo, através de discussões críticas, de rejeições e de reformulações para que sejam propostas soluções [15].

A análise e mitigação é nesse processo contínuo e recorrente de resolução em que que a atividade ganha *momentum* e uma nova estrutura emerge, dando início a novas reflexões, e assim sucessivamente. Tendo em conta estes aspetos, vamos utilizar a sequência proposta por Engeström para a sequência de passos do ciclo de aprendizagem expansiva, a qual é marcada por sete etapas (Figura 7):

- Questionamento da situação atual: onde pode ocorrer a crítica ou rejeição de alguns aspetos da prática corrente;
- Análise histórica das contradições/ análise empírica atual: esta análise da situação envolve transformações de práticas da situação em questão, para descobrir causas ou mecanismos exploratórios;
- Modelação da nova situação: implica a construção de um modelo da nova ideia que explique e ofereça uma solução para o problema;
- Exame do novo modelo: inclui a experimentação do modelo, com a intenção de perceber a sua dinâmica, potencialidades e limitações;
- Implementação do novo modelo – resulta na experimentação do modelo por meio de aplicação prática;
- Reflexão sobre o processo: inclui a avaliação do novo processo e consolidação da nova prática – significa o estabelecimento de uma nova forma de prática.

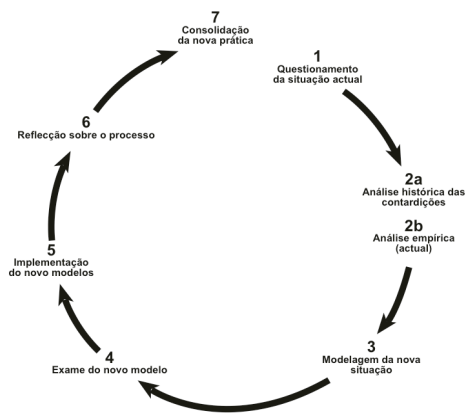

**Figura 5. Ciclo de Aprendizagem expansiva**

## 6   Modelação de um serviço LDS através da Teoria da Atividade

A partir da Teoria da Atividade é possível descrever um serviço genérico segundo a lógica dominante de serviço, como sendo a interação de uma rede de sistema de atividades que partilham o mesmo objeto (i.e. o teste, analise de vulnerabilidade, etc..), mas que possuem pontos de vista distintos, ou seja existe multivocalidade. Contudo existe colaboração, coordenação e cooperação entre os dois sistemas de atividades: a atividade do cliente (quem presta o serviço de teste de intrusão) e a atividade do fornecedor (que consome o serviço), através da partilha de ferramentas que mediam a relação do sujeito com o espaço do problema, i.e. o objeto (ver figura 8).

Por colaboração dos sistemas de atividade é entendida como um contexto em que as pessoas trabalham juntas com o mesmo objetivo, produzem uma solução final comum [16]. Para além dos objetivos comuns entre os diversos participantes, cada um deles tem os seus próprios objetivos individuais, decorrentes da sua função profissional. Assim, num atividade colaborativa, a presença de objetivos comuns não é inconciliável com a continuação de propósitos próprios para cada uma das organizações. Atingir o compromisso entre estes dois tipos de objetivos não é simples, mas é essencial para a conclusão do resultado.

O conceito de coordenação está relacionado com os resultados da conversação para a ação. No decorrer da comunicação entre as organizações envolvidas numa conversação, resultam compromissos [16]. Para garantir o cumprimento destes compromissos, através da soma dos trabalhos individuais, é necessária a coordenação das tarefas. Esta coordenação é responsável pela organização do grupo para evitar que esforços de comunicação sejam perdidos e garantir que as tarefas sejam realizadas na ordem correta, no tempo correto e cumprindo as restrições e objetivos. É através da coordenação que as pessoas executam o fluxo normal e habitual da sua interação, ou seja, é uma situação na qual as tarefas e os procedimentos necessários estão divididos em funções disjuntas.

A cooperação é um modo de interação em que as pessoas focam-se num objeto comum e partilham o objetivo da atividade coletiva, em vez de cada um se preocupar com a execução das suas funções e ações atribuídas. Dito por outras palavras, a cooperação é a realização conjunta de tarefas, feita por agentes, num espaço partilhado. Os agentes planeiam juntos, atuam em conjunto, negoceiam e possuem um objetivo partilhado. A falha de um implica a falha de todos.

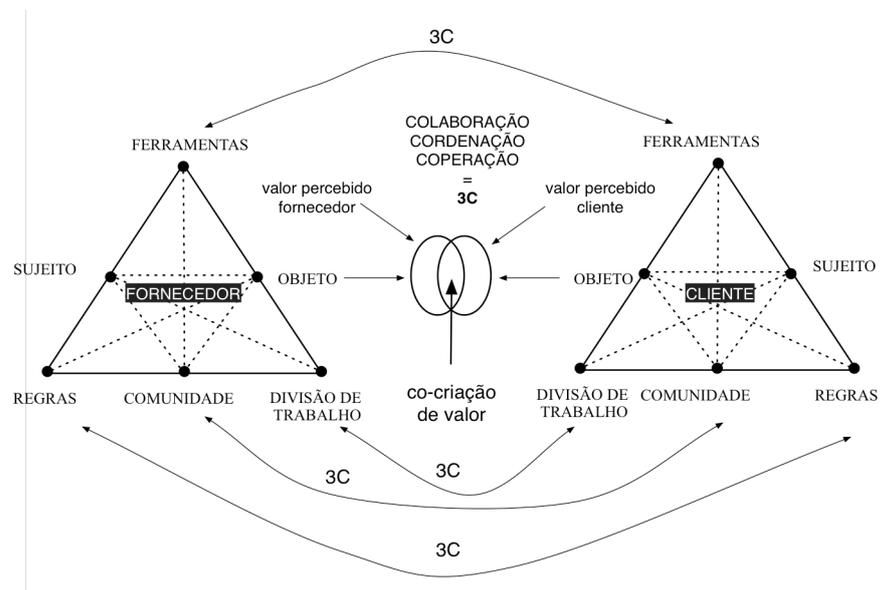

**Figura 6. Modelação da Lógica Dominante Serviço**

### 6.1 Descrição do método

A partir do modelo que descreve um serviço segundo a lógica dominante do serviço, propomos observar as manifestações das contradições (figura 6) descritas na teoria da atividade. Esta observação será feita através doe uma adaptação do ciclo de aprendizagem expansiva, tal como descrito por Engeström (figura 7), de modo a reconhecer o modo como as manifestações das contradições são mitigadas de modo a permitir manter a operação do serviço. A solução proposta possui as seguintes etapas e respetiva relação com o ciclo de aprendizagem expansiva (figura 9):

1) **Diagnóstico:** Corresponde as fases de questionamento e descoberta das tensões existentes no serviço.
    a. Analisar empiricamente o trabalho realizado na prestação do serviço;
    b. Descobrir as causas das tensões existentes e classifica-las de acordo com o tipo de contradição: 1ª, 2ª, 3ª ou 4ª ordem;

2) **Recuperação:** Corresponde a identificação das soluções encontradas de modo a manter a operação do serviço.

a. Através da interpretação das contradições identificadas a operação do serviço é alterado de modo a mitigar as tensões;
3) **Monitorização**: Corresponde a observação dos resultados das transformações realizadas e o seu efeito no serviço (cliente e/ou fornecedor).
   a. fazer o exame da nova operação do serviço com a intenção de perceber a sua dinâmica, potencialidades e limitações e o seu impacto no cliente/fornecedor;
   b. Voltar a etapa 1 (ações de diagnóstico).

O diagnóstico será baseado em entrevistas, em especial o modelo de entrevista etnográfica sob um formato de entrevistas semi-estruturadas, para possibilitar a obtenção de mais informações, ao longo da entrevista, acerca do tema em estudo. O modelo proposto será um formato composto por uma série de conversas, nas quais o entrevistador introduziu lentamente novos elementos para auxiliar os entrevistados a responderem como informantes, evitando que as entrevistas se assemelhem a um interrogatório formal, que conduziria a perda de harmonia e os informantes poderiam acabar por suspender a sua cooperação.

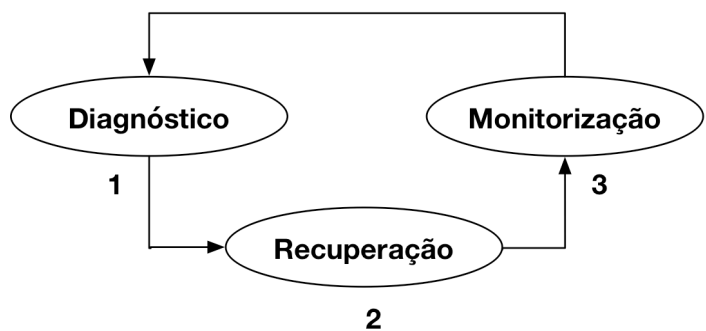

**Figura 7. Etapas do modelo.**

## 7   Caso de Estudo

A solução proposta na secção 6 foi aplicada na análise de um artefacto de suporte ao SGSI. Para o efeito, o caso de estudo foi realizado numa organização.
Foi utilizado o método proposto: Diagnóstico, Recuperação e Monitorização. As técnicas de colheita de informação usadas foram: entrevistas etnográficas, análise de documentos e conversas informais. Todos os diálogos foram analisados e serviram de base para o estudo. Os documentos e relatórios existentes foram também usados na análise através de cópias das suas imagens.
   O trabalho foi conduzido com o auxílio de 10 especialistas de segurança de informação (ver tabela 4). Os especialistas possuem um leque variado de experiência e valências profissionais em área da segurança da informação, tendo um deles experiência em cargos de chefia. Todos os especialistas estão familiarizados com artefactos de colaboração, coordenação e cooperação sobretudo para o controlo pessoal da sua atividade e partilha da informação com os clientes.

   .

**Tabela 4. Valência dos consultores.**

| CONSULTOR | EXPERIÊNCIA PROFISSIONAL | NÚMERO CONSULTORES | VALÊNCIA |
|---|---|---|---|
| Júnior | 3 Anos | 5 | Conhecimentos de análise de risco de sistemas de gestão de informação e testes de intrusão. |
| Sénior | 8 Anos | 4 | Valências na realização de testes de intrusão. |
| Coordenador | 15 Anos | 1 | Capacidade de liderança. |

No decorrer da análise das ações e operações realizadas ao longo do serviço, é possível identificar diferentes tipos de tensões. Contudo foi difícil a identificação de tensões de 1ª ordem, i.e., tensões encontradas num elemento isolado de cada uma das atividades. Em muitos dos casos as tensões surgem como manifestações entre dois elementos da atividade (i.e., contradições de 2ª ordem).

Durante o caso de estudo, verificou-se a existência de tensões entre o sujeite (i.e. o consultor, cliente) e o objeto (i.e. testes), nomeadamente o tipo de teste e o tempo despendido. Tanto o cliente como o fornecedor tinham perspetivas distintas. Ou seja, existia uma discrepância entre o serviço prestado e o serviço esperado. Este tipo de tensões eram recorrentes e resultava num dispêndio grande de temo coordenação para conseguirem chegar a um acordo.

A análise desta tensão levou-nos a observar a existência das contradições que estavam na raiz das tensões. Na tabela 5 são apresentadas as principais contradições

**Tabela 5. Contradições observadas no serviço.**

| | **Relação entre Sujeito e objeto** | **Relação entre Sujeito e Regras** | **Relação entre comunidade e objeto** |
|---|---|---|---|
| **1** | Dificuldade em definir os controlos que devem ser observados. | Não existe um modo comum de definir a observação que devem ser realizado de acordo com a vulnerabilidade encontrada. | A divisão de trabalho é feita de um modo *ad-hoc* o que dificulta o planeamento do trabalho. |
| | **Relação entre Sujeito e Ferramenta** | **Relação entre Ferramenta e Objeto** | **Divisão de trabalho** |
| **2** | Dificuldade em definir os passos feitos num ação de gestão de segurança. | Cada sujeito utiliza ferramentas distintas para executar ação de gestão de segurança. | As ações e operação realizadas não são explicitas o que dificulta a compreensão do trabalho realizado. |
| | **Relação entre objeto do cliente e objeto do fornecedor** | | |
| **3** | O cliente e o fornecedor têm dificuldade em encontrarem um entendimento do tipo de controlos que devem ser observados, ou seja o âmbito do serviço. | O cliente e o fornecedor têm dificuldade em definir o esforço (i.e. tempo, capacidade dos consultores) necessário para realizar o trabalho. | Devido às alterações constantes no cliente, o trabalho tem que ser feito em janelas temporais definidas pelo fornecedor e que corresponde a cenários em que os sistemas de informação do fornecedor tem alguma estabilidade. |

Uma vez que não seria possível resolver todas as contradições procurou-se analisar se o uso do artefacto *27001 Manager* permitiria mitigar algumas delas. Para tal optou-se por tentar mitigar as contradições existentes entre o sujeito (quem faz o teste de intrusão/quem valida o teste) e o objeto (i.e. o teste que deve ser feito). Na verdade, verificou-se que a iteração entre o sujeito e o objeto não é feita diretamente mas sim através da mediação de um artefacto, que constitui o esforço necessário para realizar o SGSI e que dependente da multivocalidade do objeto partilhado entre o cliente e o fornecedor. Procurou-se mitigar esta contradição através da normalização de um artefacto/ferramenta (figura 10).

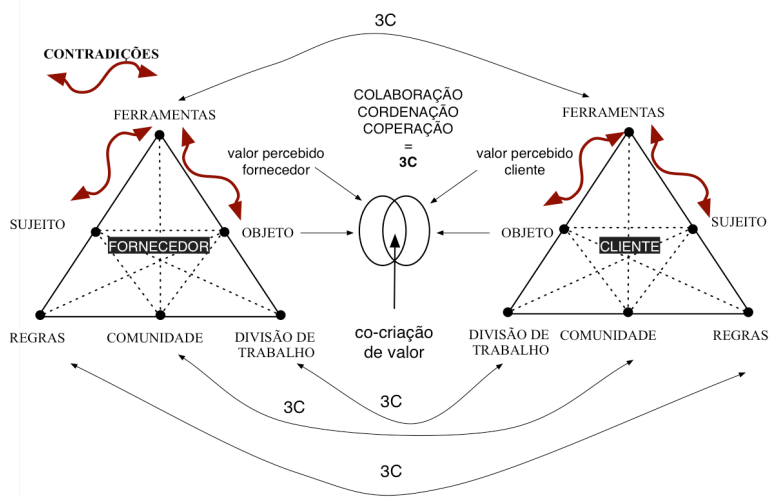

**Figura 8. Contradições Observadas**

O artefacto desenvolvido constitui um suporte e integrada, de uma grande parte as etapas associadas a este processo, nomeadamente, o levantamento e identificação de procedimentos e requisitos operacionais, a conceção, desenvolvimento e implementação do sistema em si, e ainda a fase de gestão operacional, com total controlo e capacidade de análise e *reporting*, tanto em tempo real.

Realça-se a facilidade de promover a colaboração, coordenação e cooperação entre o fornecedor e o cliente. Com o *27001 Manager* verifica-se a possibilidade de mitigar os problemas, o planeamento de longo prazo, o trabalho em equipa, a tomada de decisões fundamentada em evidências, a melhoria contínua, a estruturação organizacional horizontal e descentralizada e balizar o esforço de co-criação de valor entre o fornecedor e o cliente. Para tal tornou-se explicito a medida do esforço para realizar os testes de intrusão.

Neste sentido foi analisado e definido em conjunto com a equipa de consultores (dos clientes e fornecedores) o uso do *27001 Manager* de modo a contemplar dois aspetos: a identificação dos controlos e o esforço necessário para a sua monitorização. O resultado deste trabalho está presente na tabela 6 e figura 11 e descreve os benefícios do uso do *27001 Manager*

**Tabela 6. Benefícios do *27001 Manager*.**

| Módulo | Descrição |
|---|---|
| Risco | O apoio ao processo de definição da metodologia de análise de risco |
| Plano de ação | Definir a metodologia e consequente plano de ação para a implementação eficiente da solução. Suporta a complexidade do processo, uma vez que um sistema SGSI necessita de informação proveniente dos sistemas existentes na organização e de informação existente na memória organizacional, ou seja, nos seus colaboradores. |
| Monitor | Mantém a coerência entre os processos e registos mais concretamente: Incidentes de Segurança; Risco e Mitigação; Inventário; Alterações e Alterações |
| *Audit* | Normalização do processo |
| Portal | Possibilita a fácil comunicação entre todos os interlocutores |

O processo de utilização do *27001 Manager* engloba a necessidade de compreender o trabalho realizado pelas pessoas na organização e ao mesmo tempo constitui uma aprendizagem para que seja possível melhorar as medidas obtidas, a partir da monitorização contínua do serviço. As expectativas encontradas foram: 1) uma base de melhoria contínua do esforço e 2) estimular os consultores a normalizarem as competências apreendidas, tendo em conta a perceção que os clientes têm do trabalho realizado.

O processo de aprendizagem implica a sua estruturação, de forma consciente, para obter a informação através de um processo participativo, onde o cliente e os fornecedores cooperam. O processo em si não é neutro, uma vez que motiva mudanças de pensamentos, ações e condutas nos clientes e no próprio fornecedor.

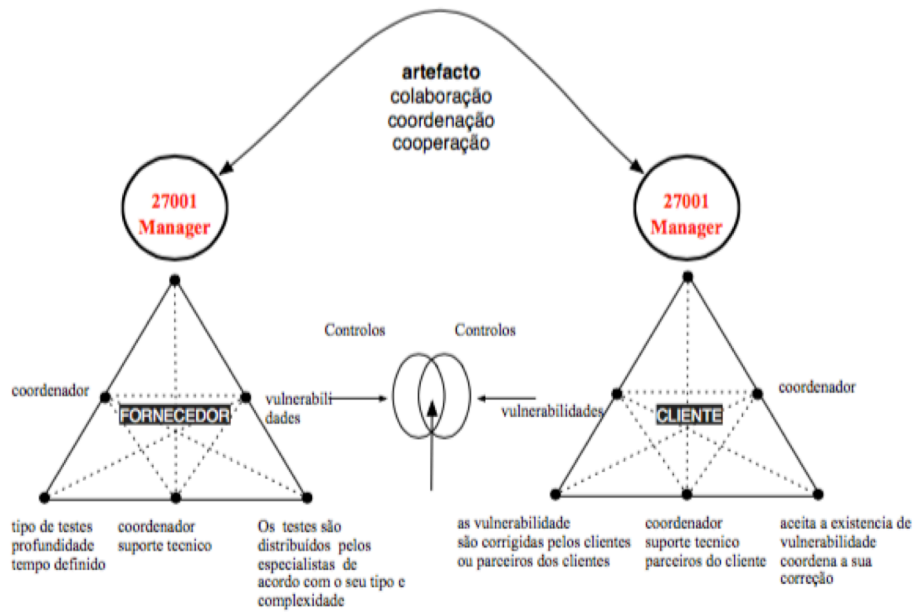

## 8  Trabalhos Relacionados

Não obstante existirem um conjunto relevante de métodos destinados ao suporte de SGSI, dos quais destacamos [17]: *COBRA Risk*, *RealISO*, *VSRisk* e Módulo *Rosk Manager*, que procuraram através de controlos, estruturação, avaliação de cenários e modelo de maturidade, nenhum deles se preocupa em definir a relação entre o cliente e fornecedor do ponto de vista de um serviço.

Na visão de Vargo e Lusch [5] os clientes pretendem, mesmo quando estão a adquiri um produto, ter um serviço. Os produtos são um meio para a realização de um serviço. Os clientes deixam de ser espectadores para serem agentes ativos no novo paradigma. Existe a necessidade da participação do cliente. Este torna-se um co-criador de valor durante o serviço.

Wangsa et al. [18] descrevem um quadro de análise de requisitos com base em vários princípios idênticos aos existentes na teoria da atividade para melhorar a análise de requisitos para a prestação de um serviço, tendo em conta os seguintes critérios: i) a necessidade de adequar as necessidades específicas de cada cliente; ii) a necessidade de lidar com a existência de conflito resultantes de necessidades distintas dos clientes, e por fim iii) a necessidade de ter em atenção a existência de um contexto dinâmico em constante transformação nas organizações.

Segundo Whitman, [6] é essencial que as organizações compreendam o contexto em que operam para que as suas abordagens de segurança da informação possam mitigar os problemas atuais e futuros. Aponta também uma solução recorrente, que para resolver um problema tecnológico, por vezes é feito através de mais tecnologia, contudo tal não conduz ao benefício esperado e traduz-se em desperdício de recursos financeiros.

Kennedy [6] refere que deve ser dada mais atenção aos erros humanos ou falhas. Isto inclui atos realizados sem propósito malicioso por um utilizador autorizado. Independentemente da causa, até mesmo erros inócuos podem produzir danos extensos. Por exemplo, um erro de digitação simples pode causar interrupções em toda a rede. Um exemplo clássico foi a interrupção em 1997, onde cerca de 45% da internet ficou inoperacional durante algumas horas.

Thakor e Kumar [19] afirmam que pode existir um desequilíbrio de conhecimentos e experiências entre o fornecedor e o cliente, o que dificulta a comunicação e cooperação entre as partes. É o caso dos serviços profissionais, uma vez que estão associados à complexidade, conhecimento especializado, e um elevado nível de incerteza que desafia tanto o fornecedor como o cliente durante a prestação do serviço. Para o fornecedor, pode ser problemático apresentar a oferta de serviço e gerir o processo de modo a obter o resultado esperado pelo cliente. Para o cliente, pode ser problemático compreender e avaliar o serviço e para identificar e integrar os seus recursos no serviço.

Mcgraw [20] denota que não obstante existirem ferramentas de suporte aos testes, adequadas para auxiliar a identificação e exploração das vulnerabilidades, a avaliação humana é necessária para revelar falhas no projeto ou vulnerabilidades ao nível de execução mais complexas, sendo esta tarefa onerosa, devendo portanto ser definida de um modo claro a relação entre os clientes e os fornecedores.

Umarao [21] observa que os testes de intrusão juntamente com a análise de vulnerabilidades são necessárias para garantir a segurança dos sistemas de informação, contudo verifica que o processo é geralmente complexo, resultando num trabalho intensivo, necessidade de existir consultores com um vasto leque de conhecimento de diferentes métodos e ferramentas pelo que conclui que é necessário a existência de métodos normalizados.

Verendel [22] ao examinar de um modo crítico vários trabalhos sobre a representação quantitativa e análise de segurança a partir de 1981 conclui que a validade dos resultados da maioria dos métodos é surpreendentemente ainda pouco clara devido a não existir uma comparação entre os métodos e dados empíricos.

Savola [23] afirma que a utilização de métricas é algo ainda bastante incipiente e enumera algumas razões para tal: (i) a segurança é muitas vezes considerado como um serviço eclético, sendo o resultado de elementos doutrinários de origens diversas que não chegam a ser articulados numa unidade sistemática e consistente. (ii) A segurança propriamente dita é um campo de investigação recente, (iii) há uma falta de partilha de dados credíveis para serem utilizados no desenvolvimento de métricas.

Ainda Savola [23] observa que análise do risco de segurança e seu impacto sobre um sistema de informação permanece na maior parte das vezes num nível bastante abstrato a partir da perspetiva do desenvolvimento. Por outro lado, um manancial de informação relacionada com a segurança está disponível, mas raramente é utilizado nas organizações.

Heinonen [24] identifica os mecanismos que contribuem para a segurança [1]: comunicação flexível, recuperação automática em caso de erro, verificação de integridade e disponibilidade, uso de métricas comuns e repositórios de medição, reutilização de métricas disponíveis e medidas relevantes para a segurança.

Segundo Engeström a Teoria da Atividade [25] é uma ferramenta analítica para compreender a complexidade do trabalho realizado pelas pessoas numa organização. Contudo a utilização desta ferramenta só será possível tendo em conta os seguintes pressupostos: i) o tempo de pesquisa deve ser longo o suficiente para se compreenderem os objetivos das atividades. Todo o comportamento e ferramentas utilizadas pelas organizações são objeto de um processo de transformação histórica e social, ao longo do tempo que deve ser compreendido; ii) a compreensão de uma atividade implica estabelecer os tipos de mudanças que poderão ocorrer nela e nas relações com outras atividades; e iii) a análise deve ser abrangente. Primeiro, devem considerar-se os padrões mais amplos da atividade, antes de considerar os fragmentos episódicos que não revelam a direção geral e a importância de uma atividade.

Os trabalhos descritos na área da segurança de informação exploram o acesso à informação indevida e sugerem abordagens sobretudo relacionadas com problemas tecnológicos, deficiências processuais ou engenharia social, mas não analisam explicitamente a existência de ferramentas que possibilite a oferta de um serviço entre o cliente e o fornecedor. Contudo realçam a necessidade de existirem mecanismos de controlo da complexidade da relação entre os clientes e os fornecedores.

Por outro lado as metodologias partilham um mesmo desígnio, voltado para o nível operacional da organização. Ou seja, têm um objetivo imediato de assegurar o nível de segurança no presente momento, não procurando responder a questão relacionado com as atividades que devem ser realizadas para obter um nível de segurança no futuro, porque normalmente não têm em atenção a mudança constante da organização.

## 9 Conclusões

Um sistema de atividade apresenta uma multiplicidade de perspetivas, tradições e interesses da comunidade do sistema. Pela divisão de trabalho, posições e pontos de vista diferentes são atribuídos aos diferentes participantes que, por sua vez, já carregam suas próprias histórias.

A análise de um serviço, segundo a lógica LDS, através de uma rede de sistema de atividades, onde existe a atividade do cliente e a atividade do fornecedor, promove a noção da multivocalidade quando se considera redes de sistemas de atividade que interagem entre si. Isto significa que promove a existência de várias perspetivas. Atividades são formações coletivas, e isso significa que nenhum participante (cliente e fornecedor), partilha exatamente a mesma visão, a mesma perspetiva, os mesmos interesses com os outros. Esta perceção pode auxiliar a identificação das tensões/contradições que estão na origem dos problemas. Permite ainda encarar os problemas como um modo de evoluir as atividades.

Isto resulta na existência de diversos aspetos da rede de sistemas de atividade relacionados com os serviços. Os serviços onde todos perdem, corresponde a existência de uma rede de sistemas de atividade muito segmentada, onde existe fragmentação ou então uma estagnação quando existe a unificação total na rede dos sistemas de atividade.

Da experiência obtida com o estudo de caso, é importante destacar que podemos a partir dos diagramas de atividades descrever um artefacto de suporte ao desenvolvimento de um SGSI. As atividades passam a ser vistas como unidades de análise, agregando uma série de informação relacionada com o trabalho realizado pelo cliente e pelo fornecedor. A atividade forma um contexto de análise onde as ações, pessoas, ferramentas e regras são identificadas neste contexto. A análise das tensões permitem identificar as contradições e assim mitigar as contradições.

A análise através da Teoria da Atividade, do artefacto modelo permitiu compreender a sua dinâmica, nomeadamente na identificação das propriedades e pormenorização de viabilidade e controlo.

Deste modo será possível, de uma forma sistemática, analisar e propor alterações no serviço, para que seja possível passar do abstrato para o concreto através de ações de aprendizagem, discussões críticas, rejeições e de reformulações, para que sejam propostas novas soluções para a participação e a interação entre o cliente e o fornecedor.

Pretende-se analisar se as soluções partir da teoria da atividade podem vaticinar uma aprendizagem para o aperfeiçoamento, e em que cenários podem condicionar a evolução.

Como conclusão, podemos afirmar que representar um serviço, a partir da teoria da atividade, constitui uma base sólida que promove a análise do trabalho realizado pelas pessoas e que em ofertas complexas, tais como os testes de intrusão, constitui uma proposta hermenêutica, formulada através da interação entre o cliente e o fornecedor, suportada na melhoria continua através da colaboração, coordenação e cooperação.